\documentclass[a4paper,conference]{IEEEtran}
\IEEEoverridecommandlockouts

\usepackage{amsmath,amssymb,amsfonts,amsthm,dsfont}
\usepackage{babel}
\usepackage{graphicx}
\usepackage{textcomp}
\usepackage{xcolor}
\def\BibTeX{{\rm B\kern-.05em{\sc i\kern-.025em b}\kern-.08em
    T\kern-.1667em\lower.7ex\hbox{E}\kern-.125emX}}

\usepackage{tikz}
\usetikzlibrary{arrows.meta, positioning, shapes.geometric, calc}

\usepackage{pgfplots}
\usepackage{pgfplotstable} 
\pgfplotsset{compat=1.18}

\definecolor{Red}{RGB}{190,30,60}
\definecolor{Black}{RGB}{0,0,0}
\definecolor{Gray}{RGB}{128,128,128}
\definecolor{LightGray}{RGB}{160,160,160}
\definecolor{White}{RGB}{255,255,255}

\definecolor{LightYellowOrange}{RGB}{255,200,42}
\definecolor{OrangeBright}{RGB}{225,109,0}
\definecolor{DarkBordeaux}{RGB}{113,28,47}

\definecolor{SkyBlue}{RGB}{102,180,211}
\definecolor{BlueLight}{RGB}{0,112,155}
\definecolor{BlueDark}{RGB}{0,63,87}

\definecolor{GreenLight}{RGB}{172,193,58}
\definecolor{GreenMid}{RGB}{109,131,0}
\definecolor{GreenDark}{RGB}{0,83,74}

\definecolor{VioletLight}{RGB}{138,48,127}
\definecolor{VioletMid}{RGB}{81,18,70}
\definecolor{VioletDark}{RGB}{76,24,48}

\usepackage{tcolorbox}

\newcounter{boxcounterR} 
\newcounter{boxcounterG} 
\newcounter{boxcounterB} 

\DeclareMathOperator{\supp}{supp}

\definecolor{darkgreen}{rgb}{0.0,0.0,0.0}
\theoremstyle{remark}
\newtheorem{remark}{Remark}
\usepackage[
  acronym,      
  nomain,       
  xindy,        
  shortcuts     
]{glossaries}
\usepackage{algorithm}
\usepackage{algpseudocode}
\usepackage{hyperref}
\makeglossaries
\newacronym{5g}{5G}{Fifth-Generation Wireless Systems}

\newacronym{ai}{AI}{Artificial Intelligence}
\newacronym{ann}{ANN}{Artificial Neural Network}
\newacronym{ask}{ASK}{amplitude-shift keying}
\newacronym[plural={AWGN}, \glsshortpluralkey={AWGN}]{awgn}{AWGN}{additive white Gaussian noise}

\newacronym{bch}{BCH}{Bose-Chaudhuri-Hocquenghem}
\newacronym{bec}{BEC}{binary erasure channel}
\newacronym{ber}{BER}{bit error rate}
\newacronym{bler}{BLER}{block error rate}
\newacronym{bpsk}{BPSK}{binary phase-shift keying}
\newacronym{bsc}{BSC}{binary symmetric channel}
\newacronym{cdf}{CDF}{cumulative distribution function}
\newacronym{club}{CLUB}{Contrastive Log-ratio Upper Bound}
\newacronym{cnbmm}{CNBMM}{Conditional Neural Bernoulli-Mixture Model}

\newacronym{cnn}{CNN}{convolutional neural network}
\newacronym{csi}{CSI}{channel state information}
\newacronym{csit}{CSI\babelhyphen{nobreak}T}{channel-state information at the transmitter}
\newacronym{csir}{CSI\babelhyphen{nobreak}R}{channel-state information at the receiver}

\newacronym{dmc}{DMC}{discrete memoryless channel}

\newacronym{ecc}{ECC}{error-correcting code}
\newacronym{ecdf}{ECDF}{empirical distribution function}
\newacronym{ee}{EE}{energy efficiency}
\newacronym{epi}{EPI}{entropy power inequality}
\newacronym{evd}{EVD}{eigenvalue decomposition}

\newacronym{ff}{FF}{feed-forward}
\newacronym{ffnn}{FF-NN}{feed-forward neural network}
\newacronym{fsk}{FSK}{frequency-shift keying}

\newacronym{gm}{GM}{Gaussian mixture}
\newacronym{gwtc}{GWTC}{Gaussian wiretap channel}

\newacronym{iid}{i.i.d.}{independent and identically distributed}
\newacronym{infonce}{ InfoNCE}{Information Noise-Contrastive Estimation}
\newacronym{il}{IL}{information leakage}

\newacronym{iot}{IoT}{internet of things}

\newacronym{kkt}{KKT}{Karush\babelhyphen{-}Kuhn\babelhyphen{-}Tucker}

\newacronym{kld}{KL divergence}{Kullback-Leibler divergence}

\newacronym{lb}{LB}{lower bound}
\newacronym{lhl}{LHL}{leftover hash lemma}
\newacronym{los}{LoS}{line-of-sight}
\newacronym{lvs}{LVS}{latent variable secrecy}

\newacronym{mac}{MAC}{multiple access channel}
\newacronym{map}{MAP}{maximum a posteriori}
\newacronym{mc}{MC}{Monte Carlo}
\newacronym{mdl}{MDL}{mode-dependent loss}
\newacronym{mi}{M.I.}{mutual information}
\newacronym{mimo}{MIMO}{multiple-input multiple-output}
\newacronym{mimome}{MIMOME}{multiple-input multiple-output and multiple antennas at eavesdropper}
\newacronym{mine}{MINE}{mutual information neural estimator}
\newacronym{miso}{MISO}{multiple-input single-output}
\newacronym{ml}{ML}{machine learning}
\newacronym{mlp}{MLP}{multi-layer perceptron}
\newacronym{mmf}{MMF}{multi-mode fiber}
\newacronym{mmwave}{mmWave}{millimeter wave}
\newacronym{mop}{MOP}{multi-objective programming problem}
\newacronym{mrc}{MRC}{maximum ratio combining}
\newacronym{msc}{MSC}{modulation and coding scheme}
\newacronym{mse}{MSE}{mean squared error}
\newacronym{mvb}{MVB}{multivariate Bernoulli }

\newacronym{nlos}{NLoS}{non-line-of-sight}
\newacronym{nn}{NN}{neural network}
\newacronym{nve}{NVE}{normalized validation error}
\newacronym{nwj}{NWJ}{Nguyen-Wainwright-Jordan bound}
\newacronym{pdf}{PDF}{probability density function}
\newacronym{pls}{PLS}{Physical layer security}
\newacronym{pmf}{PMF}{probability mass function}
\newacronym{pwc}{PWC}{polar wiretap code}

\newacronym{qam}{QAM}{quadrature amplitude modulation}

\newacronym{ra}{RA}{rearrangement algorithm}
\newacronym{rbf}{RBF}{radial basis function}
\newacronym{relu}{ReLU}{rectified linear unit}
\newacronym{rhs}{RHS}{right hand side}
\newacronym{ris}{RIS}{reconfigurable intelligent surface}
\newacronym{rnn}{RNN}{recurrent neural network}
\newacronym{rsa}{RSA}{Rivest\babelhyphen{-}Shamir\babelhyphen{-}Adleman}

\newacronym{sdm}{SDM}{space-division multiplexing}
\newacronym{sgd}{SGD}{stochastic gradient descent}
\newacronym{sgp}{SGP}{secure goodput}
\newacronym{sic}{SIC}{successive interference cancellation}
\newacronym{simo}{SIMO}{single-input multiple-output}
\newacronym{sinr}{SINR}{signal-to-interference-plus-noise ratio}
\newacronym{siso}{SISO}{single-input single-output}
\newacronym{skg}{SKG}{secret key generation}
\newacronym{snr}{SNR}{signal-to-noise ratio}
\newacronym{sop}{SOP}{secrecy outage probability}
\newacronym{svm}{SVM}{support vector machine}

\newacronym{tvd}{TVD}{total variation distance}
\newacronym{tuba}{TUBA}{total Correlation Upper Bound Approximation}

\newacronym{uav}{UAV}{unmanned aerial vehicle}
\newacronym{ub}{UB}{upper bound}
\newacronym{uhf}{UHF}{universal hash families}
\newacronym{urllc}{URLLC}{ultra-reliable low latency communication}

\newacronym{vclub}{vCLUB}{variational CLUB}
\newacronym{vd}{VD}{variational distance}

\newacronym{wtc}{WTC}{wiretap channel}
\newacronym{wrt}{w.r.t.}{with respect to}

\newacronym{zoc}{ZOC}{zero-outage capacity}
\newacronym{zosc}{ZOSC}{zero-outage secrecy capacity}
\newacronym{zosr}{ZOSR}{zero-outage secrecy rate}  
\usepackage{amsthm}
\newtheorem{definition}{Definition}
\newtheorem{theorem}{Theorem}

\newtheorem{lemma}{Lemma}
\renewcommand{\baselinestretch}{1}

\begin{document}

\title{Neural Estimation of Information Leakage for Secure Communication System Design \\
  \thanks{This work was supported in part by the German Research Foundation (DFG) PILSMOTS II.}
}


\tikzset{
  >=Stealth,
  thickarrow/.style = {->, thick},
  dashedarrow/.style = {->, thick, dashed},
  line/.style = {-, thick}
}

\tikzset{
  base/.style = {
    draw=none, fill=none,
    text centered, minimum height=1.2em, minimum width=2.5em
  }
}

\tikzset{
  block/.style = {base, text=black}
}

\tikzset{
  blockRed/.style    = {block, fill=Red!50, align=center},
  blockOrange/.style = {block, fill=OrangeBright!50, align=center},
  blockYellow/.style = {block, fill=LightYellowOrange!50, align=center},
  blockBlue/.style   = {block, fill=BlueLight!50, align=center},
  blockSky/.style    = {block, fill=SkyBlue!50, align=center},
  blockGreen/.style  = {block, fill=GreenMid!50, align=center},
  blockViolet/.style = {block, fill=VioletLight!50, align=center},
  blockDark/.style   = {block, fill=DarkBordeaux!50, align=center},
  blockGray/.style   = {block, fill=Gray!50, align=center},
  blockLightGray/.style   = {block, fill=LightGray!50, align=center},
}

\tikzset{
  input/.style    = {base, trapezium, trapezium left angle=70, trapezium right angle=110, fill=SkyBlue, text=black},
  decision/.style = {base, diamond, aspect=2, fill=GreenLight, text=black},
  terminal/.style = {base, ellipse, fill=VioletDark, text=black}
}

\tikzset{
  infoBox/.style  = {base, fill=White, draw=BlueDark, text=black},
  alertBox/.style = {base, fill=OrangeBright, draw=none, text=black}
}

\author{
  \IEEEauthorblockN{
    Darius S. Heerklotz \hspace{0.8cm}
    Ingo Schröder\hspace{0.8cm}
    Pin-Hsun Lin\hspace{0.8cm}
    Christian Deppe\hspace{0.8cm}
    Eduard A. Jorswieck
  }\\
  \vspace{-0.3cm}\IEEEauthorblockA{
    \{d.heerklotz, ingo.schroeder, p.lin, christian.deppe, e.jorswieck\}@tu-braunschweig.de \hspace{0.5cm}
  }\\
  \vspace{-0.3cm}\IEEEauthorblockA{
    Institute for Communications Technology, Technische Universität Braunschweig, Braunschweig, Germany
  }
}
\maketitle

\begin{abstract}
Underestimating the leakage can compromise \gls{pls}, while overestimating it may lead to inefficient system design. Therefore, a reliable leakage estimator is essential. Neural network-based estimators provide a data-driven way to estimate \gls{mi} without requiring full knowledge of the channel or source distributions. In this work, we aim to scale the blocklength of a wiretap code such that the estimator can still feasibly operate. We propose an improved \gls{mi} estimator based on the variational contrastive  log-ration upper bound framework, tailored for both discrete and continuous variables. By using a mixture of Bernoulli experts parameterized by neural networks, the estimator is able to quantify information leakage in communication systems, which employ complex data processing like \gls{uhf}. We further propose a method to utilize the proposed estimator to design the \gls{uhf} for a wiretap code or secret key generation design. Simulation results show that
{
prior methods significantly underestimate the \gls{mi}, particularly when using \gls{uhf} for higher blocklengths ($n\gg$16). The proposed method can scale the blocklength up to 255, and we conjecture that the design can scale well to even higher blocklengths given adequate training data and model size. Additionally, we contend that our proposed estimator and adaptive hash design framework offer a practical approach for extending \gls{pls} considerations for wiretap channels into the finite blocklength regime.

}

\end{abstract}

\section{Introduction}
{  \gls{pls} began with Claude Shannon’s work on secrecy in 1949 \cite{Shannon49}. In 1975, Wyner introduced the wiretap channel to achieve secrecy without secret keys \cite{Wyner75}. Csiszár and Körner later extended this idea to networks with multiple receivers \cite{Csiszar78}. These methods offer strong mathematical secrecy guarantees, independent of the eavesdropper’s power, and avoid complex key management—making them suitable for IoT and post-quantum security. {Secrecy in \gls{pls} is often measured by \gls{mi}, \gls{kld}, or variational distance \cite{bloch_physical-layer_2011}. Several schemes that utilize machine learning methods have been developed to estimate these quantities.} { Neural estimators of \gls{mi} commonly use Monte-Carlo samples to fit a distribution, then compute \gls{mi} via derived bounds involving the \gls{nn}, though each bound has trade-offs. The Donsker–Varadhan (DV) form used in \gls{mine}~\cite{belghazi_mutual_2018} can \emph{in principle} reach the exact \gls{mi}, yet its variance and the value of the critic explode exponentially as the true \gls{mi} grows. The linearised {\gls{nwj}/\gls{tuba}}~\cite{poole_variational_2019} bound is more stable but can still drift when data are scarce. \gls{infonce}~\cite{oord2018cpc} keeps variance low by turning the problem into a contrastive game, although they underestimate large \gls{mi} unless the batch of negative samples is huge. Upper-bound methods like \gls{club} \cite{cheng_club_2020} or Barber–Agakov \cite{agakov_im_2004} are handy for {minimising} \gls{mi} but become loose when \gls{mi} is small and demand extra models. Aditionally, all these approaches face one common difficulty: when the hidden vector observed by the eavesdropper has many components, either the critic’s outputs overflow the floating-point data type or the bound saturates, making optimisation unreliable.  
\textcolor{darkgreen}{This paper is motivated by the need for accurate leakage estimates in the short/finite blocklength regime. Second-order analysis \cite{Yury_FBL} shows that we cannot achieve zero error and a positive rate simultaneously. The additional
secrecy constraint creates a tradeoff between blocklength, secrecy rate, \gls{ber}, and leakage \cite{yang_wiretap_2019}. However, these analyses are asymptotic and may not be reliable for short blocklengths due to approximation errors. For practical design, instead, we propose the use of leakage estimators based on real measurements and mathematical bounds. In addition, the estimator can help fine-tune the system—for example, by adjusting the hash function output length in privacy amplification based wiretap coding \cite{hayashi_construction_2010} or \gls{skg}. In particular, the hash function output size can be initialized by approximating smooth min-entropy with conditional entropy plus a correction term, estimated via \gls{mi}. This enables an iterative, closed-loop design to minimize leakage while meeting reliability constraints—without requiring complex estimators of smooth min-entropy, as will be shown Sec. \ref{Sec_hash}.}
 
{ The main contributions of this work are: (1) We design an enhanced \gls{mi} estimator based on \gls{vclub} \cite{cheng2020club}, tailored for discrete variables and soft bits in communication systems using \gls{nn}-parametrized Bernoulli mixture experts. (2) We propose an efficient hash function design framework based on the proposed estimator and a derived bound of the conditional smooth min-entropy in terms of conditional entropy with a correction term in close form, which can be used for wiretap coding and \gls{skg}. (3) We investigate the finite blocklength performance of the proposed scheme, which demonstrates the proposed method can scale the blocklength up to 255, while prior methods significantly underestimate the \gls{mi}, particularly when using \gls{uhf} for higher blocklengths ($n\gg$16)}.

\section{Preliminaries and system model}
\subsection{The wiretap channel}
{ The basic building block in \gls{pls} is the discrete memoryless wiretap channel, which contains a legitimate transmitter Alice, a legitimate receiver Bob, and an eavesdropper Eve, where the system is described by $\mathcal X, \mathcal Y, \mathcal Z, P_{YZ|X}$, are the alphabets at the channel input, channel output at Bob and Eve, respectively, controlled by the distribution $P_{YZ|X}$ as shown in Fig. \ref{fig:sys}. There are two constraints: the reliability constraint: Pr$(M_s)\leq \epsilon_n$ where $M_s$ is the secure message transmitted by Alice and to be decoded by Bob, $\epsilon_n$ is a sequence converging to zero, and the secrecy constraint: a commonly adopted one is the strong secrecy measure by $I(M_s; Z^n)\leq\delta_n$, where $Z^n$ is the received sequence at Eve, $\delta_n$ is a sequence converging to zero. 

\vspace{-0.2cm} 
\subsection{Mutual information estimators}
A key challenge in neural \gls{mi} estimation is determining the required sample size for accurate estimation. In~\cite[Eq. 15]{belghazi2018mine}, the authors claim that sample complexity scales as ${O}\!\Bigl(\frac{d}{\varepsilon^2}\Bigr)$, where $d$ is the dimension of the parameter space and $\epsilon$ is the error tolerance. However, later~\cite{mcallester2020formal} showed that this claim is incorrect.  The main mistake is that it applies a Hoeffding's inequality to a function that can grow without bound, which is where the inequality becomes invalid. Hence, the number of samples might be larger— especially if the data has outliers that strongly affect \gls{mi}.  
If the estimator never sees these rare events, it might severely underestimate or overestimate the true \gls{mi}. Other studies~\cite{poole2019variational} have also found that the Donsker–Varadhan-based estimator becomes very unstable when the true \gls{mi} is large. When the \gls{mi} is small, it tends to give very weak estimates that cannot distinguish small differences.
In addition, computing \( I(X;Y) \) exactly requires access to the true conditional distribution \( P_{Y|X} \), which is usually unavailable. Moreover, even estimating this quantity from samples is known to be statistically hard in \emph{high} dimensions due to the curse of dimensionality and the need for density estimation.

While \gls{mine} suffers from high variance, instability and bias during training due to the exponential term—especially in high dimensions— \gls{club} \cite{cheng2020club} offers more stable \gls{mi} estimation due to the linear log-ratio structure after applying Jensen's inequality, as it avoids the instability caused by exponentiation during optimization. The central operation for the \gls{vclub} is to replace the true \( P_{Y|X} \) with a parameterized approximation \( q_\theta(Y|X) \), which remains an upper bound under mild conditions and is typically implemented as an \gls{nn}. This leads to the following:
\vspace{-0.1cm} 
\begin{equation}
\small
  \hat{I}_{\mathrm{vCLUB}}
  \;=\;
  {\mathbb{E}_{p(x,y)}
    \!\bigl[\log q_\theta(x\mid y)\bigr]}
  -
  {\mathbb{E}_{p(x)p(y)}
    \!\bigl[\log q_\theta(x\mid y)\bigr]}.
  \label{eq:vclub}
\end{equation}
This bound is motivated by the contrast between the expected log-likelihood of positive pairs \( (x,y) \sim P_{XY} \) and the average log-likelihood over negative samples where \( x \sim P_X \), \( y \sim P_Y \) are sampled independently. As such, it resembles the structure of contrastive learning objectives and can be optimized from sample data using stochastic gradient methods.


}

{
\subsection{Universal hash family}\label{sec:UHF}
\begin{definition}
    A finite family of functions
$  \mathcal{F}\subseteq \{f : \mathcal{X} \to \{0, ..., 2^k-1\}\},
    \quad\text{where } $ is called $c$-\emph{universal} if for all $x,y\in \mathcal{X},\,x\neq y$,
$  \Pr_{f\in\mathcal{F}}\!\bigl[f(x)=f(y)\bigr] \;=\; \frac{c}{2^k}.$
\end{definition}

The purpose of applying a uniformly selected hash function from \gls{uhf} in a data transmission system is to achieve privacy amplification, in the same way as it is used in \gls{skg}. Hayashi in \cite{hayashi_construction_2010} shows that the concatenation of a reverse-hash and hash pair as the outer processing of a point-to-point channel code can achieve the secrecy capacity with strong secrecy by applying the \gls{lhl} \cite[Corollary 7.22]{tyagi_information-theoretic_2023}:

\begin{theorem}\label{Th_LHL}
    For a given distribution $\mathsf{P}_{X Z}$ on $\mathcal{X} \times\mathcal{V} \times  \mathcal{Z}$, and for a mapping $F$ chosen uniformly at random from a UHF $\mathcal{F}$, $K = F(X),\,k\in\{0,1\}^{\ell}$, satisfies
\vspace{-0.2cm} 
\begin{align}
    d_{\mathrm{var}}\!\left(\! \mathsf{P}_{\!K V Z F},\!\!\; \mathsf{P}_{\!\mathrm{unif}}\!\! \times \!\!\mathsf{P}_{\!VZ} \!\!\times\!\! \mathsf{P}_F \!\right)\!\!
    \le\! 2\epsilon\!+\!\!2^{\frac{\ell \log|\mathcal V|- H_{\min}^{\varepsilon}(X|Z)}{2}-1},
    \label{EQ_LHL}    
\end{align}
\vspace{-0.2cm} 
where we define $Q_{XZ}$ as a subdistribution and
{\small
\begin{align}
  H_{\min}^{\varepsilon}(X\!\mid \!Z)
  \!:=\!\!\!\!\!\sup_{\substack{\,\|Q_{XZ}\!-\!P_{XZ}\|_1\le\varepsilon}}
       \Bigl[\!-\log_2\!
         \bigl(\sup_{x,z}Q_{X\mid Z}(x\mid z)\bigr)\!\Bigr].       \label{eq:def_HminSmooth}
\end{align}
}
\end{theorem}
\vspace{-0.2cm}Denote the secure message size and hash function input size by \(k\) and \(q > k\), respectively. The size of the local randomness introduced by Alice is denoted by \(b = q-k \). To construct the hash, we use a randomly selected invertible matrix \(\mathbf{A}\in \textsf{GL}(q, \mathbb{F}_2)\), where \(\textsf{GL}(q, \mathbb{F}_2)\) denotes the general linear group of \(q \times q\) invertible matrices over \(\mathbb{F}_2\). The hash function \(F: \mathbb{F}_2^q \to \mathbb{F}_2^k\) maps a length-\(q\) vector \(V\) to a length-\(k\) vector by multiplying with \(\mathbf{A}\) and projecting onto the first \(k\) components: $F(V) = \zeta_k(V \cdot \mathbf{A}),$ where \(\zeta_k(\cdot)\) denotes the projection onto the first \(k\) coordinates.
The inverse matrix \(\mathbf{A}^{-1}\) satisfies the identity \(\mathbf{A}\cdot \mathbf{A}^{-1} = \mathbf{A}^{-1} \cdot \mathbf{A}= \mathbf{I}_q\), where \(\textbf{I}_q\) is the \(q \times q\) identity matrix. We define an inverse hash function over the binary finite field \(\mathbb{F}_2\), where all operations are performed within this field. Given a secure message \(m_s \in \mathbb{F}_2^k\) and a random vector \(B \in \mathbb{F}_2^b\) padded by Alice, the full input to the inverse hash is the concatenation \((m_s \,\|\, B)\), where $||$ concatenates $m_s$ and $B$. The de-hashed output is then $
\mathcal{D}(m_s) = (m_s \,\|\, B) \cdot \mathbf{A}^{-1}$,
which is used during encoding. 
}
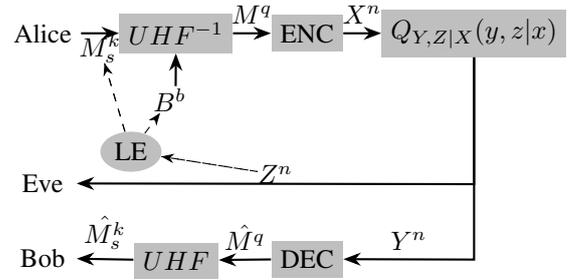
\begin{figure}[t]
    \centering
    \begin{tikzpicture}[node distance=0.5cm and 0.5cm]

  \node[block]      (alice)   {Alice};
  \node[blockGray]   [right=of alice] (UHF_inv)     {$UHF^{-1}$};
  \node[blockGray]   [right=of UHF_inv]   (BCH_enc)    {ENC};
  \node[blockGray]   [right=of BCH_enc]  (chan)   {$Q_{Y,Z|X}(y,z|x)$};

  \node[block] [below=1.5cm of alice] (eve) {Eve};

  \node[blockGray] [below=2.5cm of BCH_enc] (BCH_dec) {DEC};
  \node[blockGray] [below=2.5cm of UHF_inv] (UHF) {$UHF$};

  \node[block] [below=of eve] (bob) {Bob};
  
  \draw[thickarrow] (alice) -- node[midway, below,  inner sep=0pt, line width=0.3pt, name=MsLabel] {$M_s^k$} (UHF_inv);
  \draw[thickarrow] (UHF_inv)   -- node[midway, above] {$M^q$} (BCH_enc);
  \draw[thickarrow] ++(1.75,-0.7) coordinate (temp0) -- node[midway, below left=7pt and -4pt,  inner sep=0pt, line width=0.3pt, name=BbLabel] {$B^b$} (UHF_inv);
  \draw[thickarrow] (BCH_enc)  -- node[midway, above] {$X^n$} (chan);
  \draw[thickarrow] (chan) -- ++(0,-2cm) coordinate (temp1) -- node[midway, above,  inner sep=0pt, line width=0.3pt, name=ZnLabel] {$Z^n$} (eve);
  \draw[thickarrow] (chan) -- ++(0,-3cm) coordinate (temp2) -- node[midway, above] {$Y^n$} (BCH_dec);
  \draw[thickarrow] (BCH_dec)  -- node[midway, above] {$\hat{M^q}$} (UHF);
  \draw[thickarrow] (UHF)  -- node[midway, above] {$\hat{M_s^k}$} (bob);

  \path (MsLabel) -- (ZnLabel) -- ++(-1.9,0.3) coordinate (LeakPos);
    
  \node[fill=Gray!50, ellipse, inner sep=2pt, align=center] (leak) at (LeakPos) {LE};
    
  \draw[dashed] (MsLabel) -- (leak) -- (ZnLabel);
    
\draw[dashed,->] (ZnLabel) -- (leak);                 
\draw[dashed,->] (leak) -- (MsLabel); 
\draw[dashed,->] (leak) -- ($(leak)!0.75!(BbLabel)$);

\end{tikzpicture}
    \caption{The considered system model.}
    \label{fig:sys}
\end{figure}
\subsection{System model}
{We consider a secure communication system that employs a concatenation of a reverse \gls{uhf} and forward error correction, with transmission over a wiretap channel following \cite{hayashi_construction_2010} as shown in Fig. \ref{fig:sys}, where $M_s \sim\mbox{Unif}(\{0,1\}^k)$ is the secret message, $B \sim\mbox{Unif}(\{0,1\}^b)$ is a vector of random bits generated locally, $M \in \{0,1\}^q$ is the output of the inverse universal hash function, $q = k + b$, $X \in \{0,1\}^n$ is the encoded codeword, $Y \in \{0,1\}^n$ and $Z \in \{0,1\}^n$ are the received vectors at Bob and Eve, respectively. Bob first uses a decoder to retrieve $\hat{M} \in \{0,1\}^q$ and then applies the \gls{uhf} to retrieve the secure message $\hat{M_s} \in \{0,1\}^k$.

In our implementation, the (inverse) \gls{uhf} follows the constuction specified in Sec.~\ref{sec:UHF} and the ENC block corresponds to either a \gls{bch} \cite{bose1960} or a polar \cite{arikan2009} code. We consider either a \gls{bsc} or an \gls{awgn} channel. In the latter case, we include a simple \gls{bpsk} modulator/demodulator in the ENC/DEC blocks. The whole processing at Alice can be parameterized by $(n, k, b)$. Besides, due to the same marginal property of the wiretap channel, we can analyze it as two independent point-to-point channels.
In the following, we will develop a leakage estimator (LE block) to measure $I(M_s^k;Z^n)$, and discuss how to use it to parameterize the hash function.
}
{\subsection{Multivariate Bernoulli} \label{sec:bernoulli}
In a digital communication system with a binary alphabet, the \gls{mvb} could model the joint probability of the input $m\in\{0,1\}^k$ and output $y\in\{0,1\}^n$ where $d = k + n$, note that the channel output $y$ can be non i.i.d. 
Let $v:=(m||y)\in\{0,1\}^d$. The \gls{mvb} is parameterized in terms of $2^d-1$ weights $\{0<\vartheta_{\mathbf{s}}\leq 1\}$ over all $\mathbf{s}\in\{0,1\}^d$ with definition $0^0 := 1$ as:
{\small
\begin{equation}
P(v)
=\sum_{\mathbf{s}\in\{0,1\}^d}
  \vartheta_{\mathbf{s}}
  \prod_{i=1}^d
    v_i^{s_i}(1 - v_i)^{1 - s_i},
\quad
\sum_{\mathbf{s}}\vartheta_{\mathbf{s}} = 1.
\end{equation}
}
Effectively, this formulation of the \gls{mvb} selects the corresponding $\vartheta_s$ by matching the $v$ to its $s$ in the whole space $\{0,1\}^d$, which can be viewed as a look-up table for $\vartheta_s$ for a given binary vector $v$. Another useful definition of the \gls{mvb} can be found in \cite{dai_multivariate_2012}, which describes the \gls{mvb} in its log-linear form and models dependencies across dimension by interaction terms. 
Because there are as many free parameters as all possible values in $\{0,1\}^d$, the multivariate Bernoulli is universal for modeling any $d$-dimensional discrete PMF, whereas mixtures of only $M\in\mathbb{N}$ independent Bernoullis
{\small
\begin{equation}
\Tilde{P}(v)
=\sum_{j=1}^M
  \pi_j
  \prod_{i=1}^d
    p_{j,i}^{v_i}(1 - p_{j,i})^{1 - v_i}, \quad
\sum_{j}\pi_j = 1,
\end{equation}
}
where $p_{j,i}:=P(v_i=1|j)$, $j\in\{1,\ldots,M\}$ is the mixture component index, which trade universality ($2^d$ degrees of freedom) for tractability and computability ($M(d+1)-1$ parameters).

\section{Main contributions}

This section introduces the \gls{cnbmm} architecture, designed to model conditional distributions in communication systems. The \gls{cnbmm} enables estimation of information leakage and mutual information using the \gls{vclub} bound. Additionally, we propose a systematic approach for designing and selecting UHF parameters that leverages this estimator.

{ 
\subsection{Conditional Neural Bernoulli-Mixture Model}\label{sec:cbm_model}
Based on the ability of the \gls{mvb} to model distributions in binary discrete sets, as well as the  success of mixture of experts to model complex distributions~\cite{jacobs_adaptive_1991,chen_towards_2022,dai_deepseekmoe_2024}, we approximate $p(M_s=m_s\!\mid\!Z^n=z^n)$ by a mixture of Bernoulli experts parametrized by neural networks:
{\small
\vspace{-0.1cm}\begin{equation}
q_\theta(m_s \mid z^n)
   = \sum_{e=1}^{M}
     \pi_e(z^n;\theta_g)\;
     \prod_{i=1}^{k}
     \mathrm{Bern}\!\bigl(m_{s,i}; p_{e,i}(z^n;\theta_e)\bigr),
\label{eq:mixture}
\end{equation}
}
where the \gls{cnbmm} decomposes into a  gating network with learnable parameters $\theta_g$
that produces the mixture weights $\pi_e$ and a family of
 {expert networks} with learnable parameters $\theta_e$ that output per-bit probabilities $p_{e,i},\,\forall\,i$. The weights $\pi_e$ for each expert are calculated by a \gls{mlp} with layer-normalization and ReLU activation. This \gls{mlp} maps the received $z^n$ to normalized logits $g_e(z^n)\!\in\!\mathbb R^{K}$.
Incorporating a softmax function scaled by a temperature parameter $\tau$ helps prevent premature component collapse during early stages of the training~\cite{nguyen_is_2024}:
\vspace{-0.1cm}
{\small
\begin{equation}
\pi_e(z^n;\theta_g)
= \frac{\exp\bigl(g_e(z^n)/\tau\bigr)}%
       {\sum_{j=1}^{N}\exp\bigl(g_j(z^n)\tau\bigr)},
\qquad
\tau>0.
\label{eq:gating}
\end{equation}
}
Each expert $e$ emits a $k$-dimensional logit vector
$\boldsymbol\ell_e(z^n)$, realized by
\vspace{-0.2cm}
{\small
\begin{equation}
\boldsymbol\ell_e(z^n)
 = f_e(z^n)
 + \sum_{r=1}^{R}
   w_{e,r}(z^n)\,
   \mathbf u_r \odot \mathbf v_r
 + r_e(z^n),
\label{eq:expert_logits}
\end{equation}
}
where $f_e$ is an \gls{mlp} with the same architecture as $g_e$, $\{\mathbf u_r,\mathbf v_r\}_{r=1}^{R}\!\subset\!\mathbb R^{K}$ are shared factors between experts, $w_{e,r}(z) : \{0,1\}^{N}\!\!\rightarrow\!\mathbb R$ are weight factors per sample, $r_e$ is an linear residual path similar to \cite{he_deep_2016}.
The Bernoulli parameters follow via the sigmoid:
{\small
$p_{e,i}(z^n)=\sigma\!\bigl(\boldsymbol\ell_{e,i}(z^n)\bigr)
            =\frac{1}{1+\exp\!\bigl(-\boldsymbol\ell_{e,i}(z^n)\bigr)}$}.\\
Since $z$ is not restricted to discrete values, the network can implicitly utilize soft-bit information from continuous channels such as \gls{awgn}. Then the overall parameters of the \gls{nn} model are: 
{\small
\begin{equation}
\theta
\!=\! \bigl\{\,\!
  \underbrace{\theta_g}_{\substack{\text{Gating network}\\\text{for }\pi_E(z^n)}},\!
  \quad\!
  \underbrace{\{\theta_e\}_{e=1}^M}_{\substack{\text{Expert \gls{mlp}s }f_e\text{, }r_e\\\text{and weight factors }w_{e}}}
  \!\quad\!
  \underbrace{\{\mathbf u_r,\mathbf v_r\}_{r=1}^R}_{\text{Shared low-rank factors}}
\!\bigr\}.
\end{equation}
}



\vspace{-0.1cm}
\paragraph{Training Objective}
From \gls{club} \cite{cheng_club_2020}, we follow the same surrogate as minimizing the negative log-likelihood of positive pairs. Our experiments, however, revealed that as dimensionality grows, the optimization becomes susceptible to either vanishing or highly unstable gradients of the negative log-likelihood loss as proposed in \gls{club}. To address this, we adjust the loss function and introduce regularization into the training pipeline to improve stability.
The \gls{cnbmm} parameters $\theta$ are obtained by minimizing the composite loss:\vspace{-0.1cm}
{\small
\begin{align}\label{eq:loss_function}
\mathcal{L}(\theta)=
 & \!-\frac{1}{n}\!\sum_{j=1}^{n}\log q_\theta\!\bigl(m_s^{(j)}\!\!\mid\!\! (z^n)^{(j)}\bigr)  + \lambda_{\text{div}}\!\!\!\!\!\!\!\!\!\!\!\sum_{\substack{s<t\\ s,t\in\{1,\dots,M\}}}\!\!\!\!\!\!\!\!\!\!\frac{\langle p_s, p_t\rangle}{\parallel\langle p_s, p_t\rangle\parallel_2} \notag\\
 & + \lambda_{\text{int}}\sum_{r=1}^{R}\bigl(\lVert\mathbf u_r\rVert_2+\lVert\mathbf v_r\rVert_2\bigr),\vspace{-0.1cm}
\end{align}
}
where the first term is the average negative log-likelihood, the second encourages diversity among experts via cosine similarity of their per-bit probability vectors $p_{s}(z^n):=\{p_{s,i}(z^n)\},\,s\in\{1,\ldots,M\}$, the third applies $L_2$ regularization to the shared interaction factors $\{\mathbf u_r,\mathbf v_r\}$. Scalar weights $\lambda_{\text{div}},\lambda_{\text{int}}\geq 0$ balance the contributions.
}
\vspace{-0.2cm}
{\subsection{Design of the hash function}\label{Sec_hash}
Up to now, the design of the hash function, which is not trivial, is missing. However, we can resort to the proposed leakage estimator to achieve the design goal, to provide us with a low complexity design scheme for the hash function. From the \gls{lhl} we know that the size of the hash output is a function of the collision-/min-/smooth min-entropy, while min-entropy can be upper bounded by Shannon's entropy. \gls{club}, or its improved variants such as the proposed one, can be adapted to estimate the conditional entropy with a simple modification. Note that the \gls{lhl} as shown in Theorem \ref{EQ_LHL}, while the conditioned term is continuous but not discrete, can be specialized from the quantum counterpart, e.g., \cite[Corollary 5.6.1]{renner2005security}. However, bounding the smooth min conditional entropy by conditional entropy with a correction term in closed-form, whose conditional term is continuous, seems unknown in the literature, from the authors' knowledge. In contrast, one common result as \cite[Theorem 7.25]{tyagi_information-theoretic_2023} assumes a discrete conditional term. Therefore, we aim to bridge this gap, such that the proposed \gls{mi} estimator can not only be used for leakage estimation, but also for the hash function design. 
\begin{lemma}\label{Lemma_Hmin_H_gap}
Let $X$ be a discrete random variable, $Z\in\mathcal Z$ a continuous random variable, and fix $\varepsilon\in(0,1)$.
Select a measurable set $\mathcal E\subseteq\mathcal Z$
such that $   P_Z(\mathcal E)=1-\varepsilon.$ Assume that for every $z\in\mathcal E$ the conditional \gls{pmf}
$p_{X\mid Z}(\cdot\mid z)$ satisfies
{\small
\begin{align}
   v_{z}\!:=\!\bigl|\supp_x p_{X\mid Z=z}\bigr|\!<\!\infty,
   \,
   t_{z}:=\max_{x}p_{X\mid Z}(x\mid z)\!<\!\infty.
\label{EQ_Def_Vz_Tz}
\end{align}
}
Define the random variables

\(V:=v_{Z}\mathds{1}_{\mathcal E}\) and
\(T:=t_{Z}\mathds{1}_{\mathcal E}\). With
$  \psi_{v}(t):=
       H_{\mathrm b}(t)+(1-t)\log_{2}(v-1)+\log_{2}t,$ we have
{\small
\begin{align}\label{EQ_Hmin_H_gap}
-H_{\min}^{\varepsilon}(X\mid Z)
 \;\le\;&
   -H(X\mid Z)
   +\mathbb{E}_Z\!\bigl[\psi_{V}(T)\bigr]\notag\\
   &-\log_{2}(1-\varepsilon)
   +\frac{\varepsilon}{1-\varepsilon}\,H_{\max}(X),
\end{align}}
where $\psi_{v_z}(t) \;:=\; H_b(t) + (1-t)\log_2(v_z-1) + \log_2 t,$
$H_b$ is the binary entropy, for integers $v_z\ge 2$ and $t\in[1/v_z,1]$.
\end{lemma}
The proof is relegated to Appendix {and sketched as follows.} We first consider a high-probability region \(\mathcal A_r\) for Eve’s observation \(Z\), whose complement has probability at most \(\varepsilon\). Within this region, each \(p_{X\mid Z=z}\) has a bounded support size \(m\) and maximum value \(t\). Given $z$, we can easily bound  \(H(p)-H_{\min}(p)\le \psi_m(t)\), where \(\psi_m(t)\) is a function capturing the gap between the two entropies. Taking expectations over \(z\in\mathcal A_r\) \gls{wrt} to the gap, we can bound $H_{\min}(X\mid Z,\mathcal A_r)$ by $H(X\mid Z,\mathcal A_r)$. The definition of \(\mathcal A_r\) splits $P_{XZ}$ into two parts, which further bounds $H_{\min}(X\mid Z)$ by $H_{\min}(X\mid Z,\mathcal E)$ and also bounds $H(X\mid Z,\mathcal E)$ by $H(X\mid Z)$. Finally, the definition of \(\varepsilon\)-smooth min-entropy allows one to lower-bound \(H_{\min}^\varepsilon(X\mid Z)\) by \(H_{\min}(X\mid Z,\mathcal E)\). Putting all parts together yields the desired upper bound on \(H_{\min}^\varepsilon(X\mid Z)\) in terms of the conditional entropy \(H(X\mid Z)\), with an additive penalty.

\begin{remark}
Because \eqref{EQ_Def_Vz_Tz} guarantees
\(2\le m=v_{z}<\infty\) and \(t=t_{z}\in[1/m,1)\),
all logarithmic terms that follow are finite. In particular, finite \(m\) is needed in
\eqref{EQ_schur_convex} for \((1-t)\log_2(m-1)\),
while \(t<1\) ensures that \(-\log_2 t\) and thus
\(\psi_m(t)\) in
\eqref{EQ_Hmin_UB_p} remain finite.
These bounds make \(\psi_V(T)\) an integrable random variable,
so the expectation in \eqref{eq:2a} exists, and they also imply
\(H_{\max}(X)=\log_2\!\sup_z v_z<\infty\), which is used later in
\eqref{EQ_cond_H_cond_H}.
\end{remark}

To invoke Lemma \ref{Lemma_Hmin_H_gap} and Theorem \ref{Th_LHL}, we replace $X$ by the input of the channel encoder, namely, $M$, and replace $Z$ by $Z^n$, the observed vector at Eve. Then we can substitute \eqref{EQ_Hmin_H_gap} into \eqref{EQ_LHL}, and after simple algebra, we can design $k$ as follows:
{\small
\begin{align}\label{EQ_k_conditional_H}
    k<H(M|Z^n)-g,
\end{align}
}
where we define $g:=\mathds{E}[\psi_{V}(T)]
   -\log_{2}(1-\varepsilon)
   +\frac{\varepsilon}{1-\varepsilon}\,H_{\max}(X).$
In such a way, we can use $H(M|Z^n)$, instead of $H_{min}^{\varepsilon}(M|Z^n)$, to design the hash function, where a lower bound of $H(M|Z^n)$ can be simply derived from the proposed \gls{mi} estimator together with the definition of \gls{mi}.

In the following, we discuss how to tighten the upper bound of \eqref{EQ_Hmin_H_gap}. Define the right–hand side of \eqref{EQ_Hmin_H_gap} as below:
{\small
\begin{equation}
   B(\varepsilon)
   \!:= \!(\!1-\varepsilon)\,\bar\psi(\varepsilon)
      -\log_2(\!1-\varepsilon)
      +\frac{\varepsilon}{1-\varepsilon}\,H_{\max}(X),
   \!\,0\!<\!\varepsilon\!<\!\!1,
   \label{eq:B_eps_def}
\end{equation}
}
where $\bar\psi(\varepsilon)      := \mathbb E\!\bigl[\psi_{V}(T)\,\big\lvert\,Z\in\mathcal E\bigr]$. We select \(\mathcal A_r=[-r,r]^{\,n}\)
and make it depend on \(\varepsilon\) through the AWGN
tail equation $P_Z\bigl(\mathcal A_r^{\mathrm c}\bigr)
      = 2n\,Q\!\bigl((r-1)/\sigma\bigr)
      = \varepsilon$, i.e., $r(\varepsilon)=1+\sigma\,Q^{-1}\!\bigl(\varepsilon/2n\bigr),$
then set \(\mathcal E := \mathcal A_{r(\varepsilon)}\), whence
\(P_Z(\mathcal E)=1-\varepsilon\). Inside \(\mathcal E\) we have
\[
   V \!\le\! v_{\max}(\varepsilon)\!:=\!(2r(\varepsilon))^{n},
   \mbox{ and }
   T \!\le\! t_{\max}\! :=\! 2^{-n}(2\pi\sigma^2)^{-n/2}.
\]
We can easily check that \(\psi_{v_z}(t)\) is smooth and uni-modal in the considered channels, which has a unique minimiser $   \varepsilon^{\star} := \arg\min_{0<\varepsilon<1} B(\varepsilon).$ Because the problem is one-dimensional and well-behaved,
\(\varepsilon^{\star}\) is obtained either by
solving \(B'(\varepsilon)=0\) with a few Newton iterations
or by an elementary grid/bisection search. Inserting \(\varepsilon^{\star}\) (hence
\(r(\varepsilon^{\star})\) and
\(v_{\max}(\varepsilon^{\star})\))
back into \eqref{EQ_Hmin_H_gap} yields the tightest constant
for given block-length \(n\), noise variance \(\sigma^{2}\),
and source alphabet size \(\lvert\supp P_X\rvert\). To derive the gap caused by $\mathbb{E}_{Z}\big[\psi_{{V}(T})\big]$, we first derive the $P(M|Z^n)$, and recall $M$ is the channel encoder input.  For \gls{bsc} and the \gls{awgn} channel, $P_{M \mid Z^n}(m \mid z^n)$ can be simply derived. Then we can calculate $\mathbb{E}_{Z}\big[\psi_{V}(T)\big]$ numerically with complexity $\mathcal O(N\cdot n)$, where $N$ is the number of samples used in the Monte-Carlo simulation.

In the following we propose a close-loop design for the hash function, which is summarized in Algorithm \ref{Alg1}. Specifically, one may first use a \gls{club}-based estimator to compute an initial estimate of the conditional entropy for the hash design \eqref{EQ_k_conditional_H}. Then, iteratively change the hash output size $k$ by 1-bit in each iteration until the estimated information leakage just falls below a given threshold, where in line 14 of Algorithm \ref{Alg1} the sign change tells us that either the $j-1$-th or the $j$-th round has the desired result. This approach is justified by the fact that, although \gls{club} is designed for \gls{mi} estimation rather than conditional or min-entropy, it provides a valid reference point for tuning. Ultimately, the key design objective for \gls{pls} is not the min-entropy per se, but minimizing the actual information leakage. The proposed steps provide a reference $k$ to start with, which can highly reduce the search complexity compared to an exhaustive search or resorting to a highly precise smooth min entropy estimator. In particular, in the former case, each round of search may require a complete \gls{club} computation, and the number of rounds depends on the channel distribution. In contrast, the proposed scheme may only require a few rounds around the reference point, depending on the precision of the estimated reference point. 

{
\small
\begin{algorithm}
\caption{Adaptive hash output-size design using \gls{club}-based leakage estimation}\label{Alg1}
\begin{algorithmic}[1]
\State \textbf{Input:} Maximum tolerated leakage $\varepsilon$, \gls{ecc} with blocklength $n$ and input size $q$
\State Use the proposed estimator to estimate $H(M|Z^n)$
\State Design a universal hash function $F_{k_0}: \mathcal{X} \to \{0,1\}^{k_0}$ from \eqref{EQ_k_conditional_H}.
\State \textbf{Initialize:} $k \gets k_0$
\Repeat
    \State Estimate leakage: $\widehat{I}^{(j)}(M_s;Z^n)$
    \If{$\widehat{I}^{(j)}(M_s;Z^n) < \varepsilon$}
        \State Increase hash output-size: $k \gets k + 1$
    \ElsIf{$\widehat{I}^{(j)}(M_s;Z^n) > \varepsilon$}
        \State Increase hash output-size: $k \gets k - 1$
    \Else
        \State Stop
    \EndIf
\Until{sgn$(\widehat{I}^{(j)}(M_s;Z^n)-\epsilon)$sgn$(\widehat{I}^{(j-1)}(M_s;Z^n)-\epsilon)\leq 0$}
\State \textbf{Output:} Final hash length $k$, final leakage estimate $\widehat{I}(M_s;Z^n)\leq \epsilon$\end{algorithmic}
\end{algorithm}
}
}

\section{Numerical simulation}

\begin{figure*}[t]
    \centering
     \resizebox{\textwidth}{!}{\includegraphics{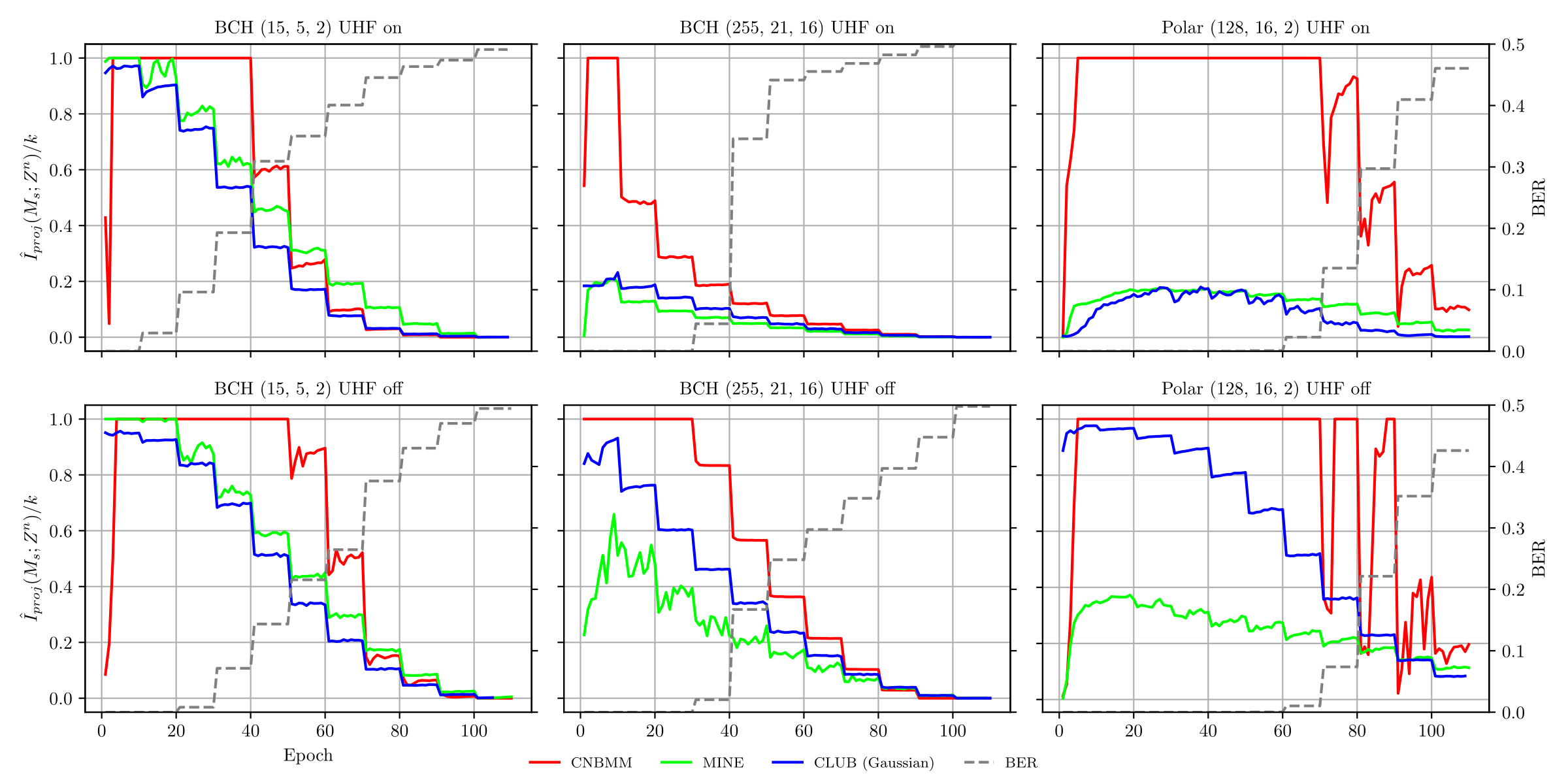}}
    \caption{Comparison of leakage estimation and BER by different schemes and setting.
    }
    \label{fig:mi_vs_snr0}
\end{figure*}
\subsection{Training Setup} \label{sec:training-setup}
{
For leakage estimation, we generate $10^6$ uniformly distributed random messages and process them through the system as shown in Fig. \ref{fig:sys}. We use BSC or BPSK + AWGN for the channel. Additionally, the channel outputs are fed through the corresponding decoder pipeline and \gls{ber} is measured, as showcased in Fig.~\ref{fig:mi_vs_ber}. We simulate each setting with and without \gls{uhf}. The estimators are trained on these samples, with leakage estimated after each epoch using an independent test set of $20,000$ samples. For the AWGN case, we start with SNR of $10$dB and reduce it by $2$dB every 10 epochs. For the BSC case, we start with $P_e = 0$ and increase it by $0.05$ every 10 epochs. We train for 110 epochs in total. The resulting \gls{mi} estimates of the \gls{cnbmm} are shown in Fig.~\ref{fig:mi_vs_snr}. 

We implement the \gls{cnbmm} as described in Sec.~\ref{sec:cbm_model} by PyTorch~\cite{ansel_pytorch_2024}.
The \gls{cnbmm} consists of $N = 2$ Bernoulli experts. The gating MLP consists of hidden dimensions $[512, 256, 12]$ with $\tau = 10$, while each expert has hidden dimensions $[2048, 512, 2048, 1024, 512, 128]$ with $R = 2n$ shared factor pairs, regularized with $\lambda_{\text{int}} = 10^{-6}$. Expert diversity is regularized with $\lambda_{\text{div}} = 0.01$. Residual connections are applied between layers that have matching hidden dimensions. 
We implement MINE as a \gls{nn} modeling the critic using six fully connected layers: expanding from 2048 to 4096 and 8192 units, then contracting to 1024 and 512 with ReLU activation, before a final linear scalar output. For the original vCLUB we parametrize $\mu$ and $\Sigma$ of a multivariate Gaussian using the same \gls{nn} as for MINE, with the exception of the last linear layer with $n$ nodes, similar to \cite{cheng_club_2020}. 
For \gls{bch} and Polar (SCL with 8 rounds of decoding) \gls{ecc} implementations, we utilize the HermesPy library~\cite{adler_hermespy_2022}, which is itself a Python wrapper around the AFF3CT project~\cite{cassagne_aff3ct_2019}. The \gls{uhf} is implemented following the methodology presented in Sec. \ref{sec:UHF}. The random seed is consistently set to 42 across all backends.
For MINE, we use a learning rate of $10^{-5}$, due to unstable training, for all other experiments we use $10^{-3}$, weight decay of $10^{-9}$, EMA decay of $0.999$ and gradient clipping with a norm of $5$ and AdaM optimizer. 

Empirically, we observed that the \gls{vclub} estimator \eqref{eq:vclub} occasionally violates the information-theoretic limit $H(M_s)=k \text{ bits}$, consistent with the observation in \cite[Sec. 3.2]{cheng_club_2020} that \gls{vclub} ceases to be an upper bound when the variational conditional differs from the true conditional. Hence, we project the raw \gls{mi} estimate $\hat{I}$ onto the interval $[0,k]$ $\hat{I}_{\text{proj}}=\min\{\hat{I}_{\text{}},k\}~\text{(bits).}$
}

\subsection{Results and discussions}
\label{sub:results_discussion}

\begin{figure}
     \centering
     \resizebox{0.5\textwidth}{!}{\includegraphics{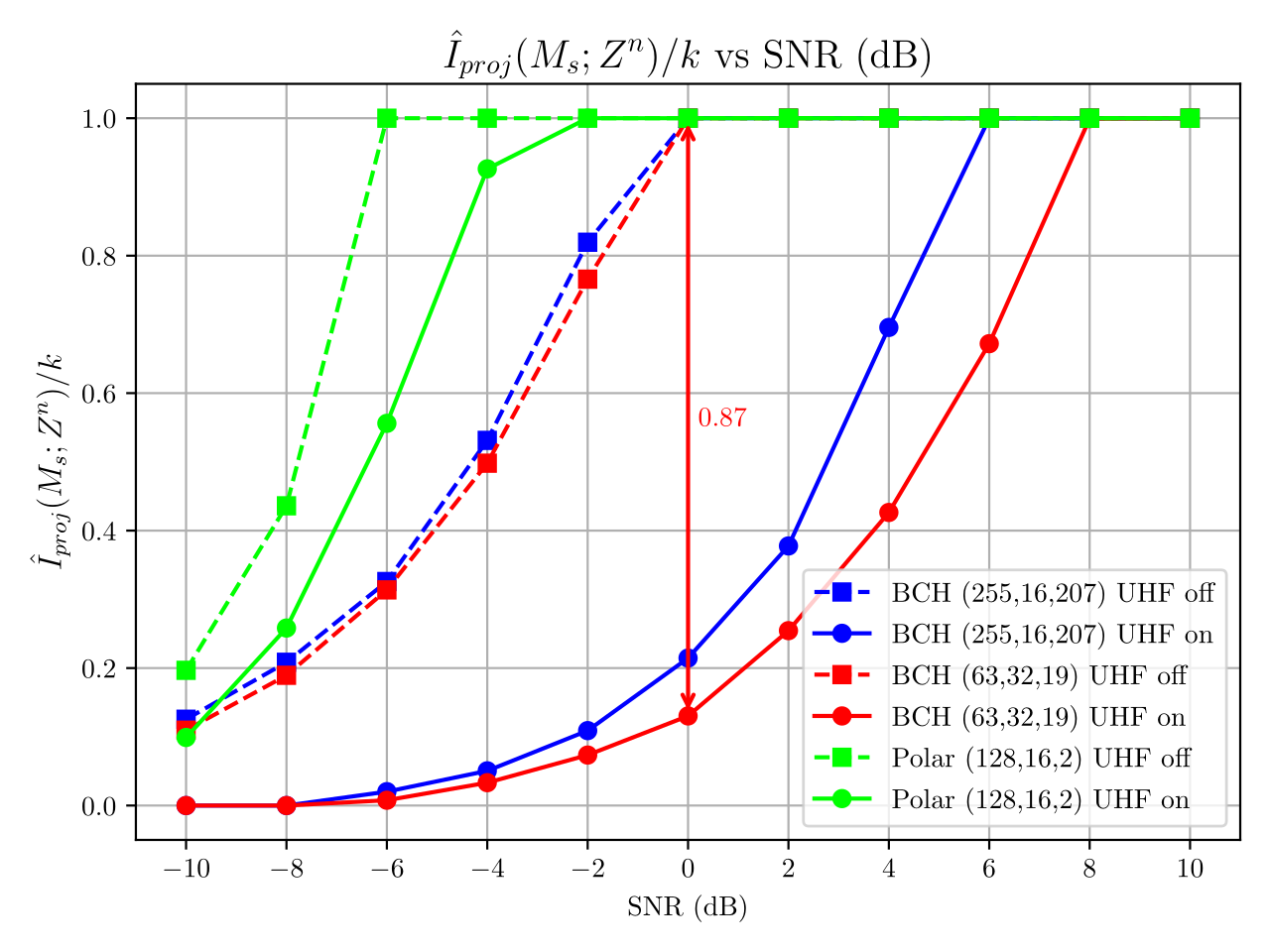}}
    \caption{Comparison of leakage estimation per information bit versus SNR for different configurations of $n, k, b$ with a \gls{bch} \gls{ecc} using the \gls{cnbmm} architecture. }
    \label{fig:mi_vs_snr}
\end{figure}

    

\begin{figure}
    \centering
     \resizebox{0.5\textwidth}{!}{
     \includegraphics{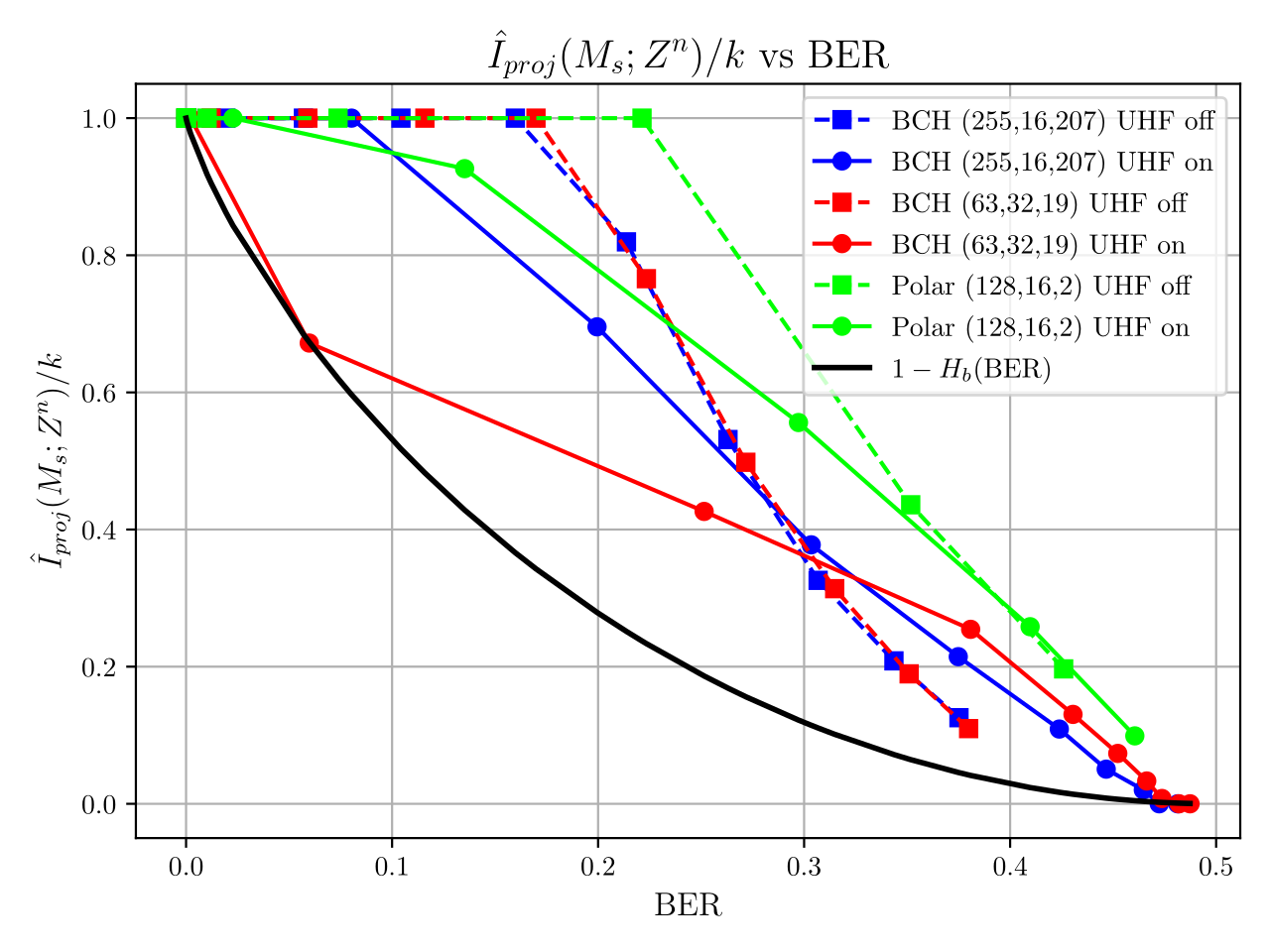}
    }    
    \caption{Comparison of leakage estimation per information bit versus BER. }
    \label{fig:mi_vs_ber}
\end{figure}

{

Fig.~\ref{fig:mi_vs_snr0} shows the normalized mutual information and BER performance over training epochs for three code: BCH (15,5,2) over BSC (left), BCH (255,21,16) over BSC (middle), and Polar (128,16,2) over AWGN (right). Top row displays results with UHF enabled, while bottom row shows performance with UHF disabled. The proposed CNBMM method (red) generally achieves higher mutual information estimates compared to conventional vCLUB + Gaussian (blue) and MINE (green) estimators, particularly in scenarios where decoding achieves low BER (gray dashed line). These results show the superior capability of the proposed method to accurately estimate information leakage in finite blocklength regime, where prior estimators tend to underestimate mutual information. For specific details regarding the systematic variation of SNR and $P_e$ across epochs for the respective channels, readers are directed to Sec. \ref{sec:training-setup}.

In Fig.~\ref{fig:mi_vs_snr} we plot $\hat I_{\text{proj}}(M_s;Z^n)/k$ versus the eavesdroppers SNR. Three codes are shown: $\mathrm{BCH}(255,16,207)$ ($R=0.87$, blue), $\mathrm{BCH}(63,32,19)$ ($R=0.81$, red) and $\mathrm{Polar}(128,16,2)$ ($R=0.14$, green) each with (circle markers) and without (square markers) \gls{uhf}. Without hashing, each code rapidly approaches full leakage ($\hat I_{\text{proj}}\approx1$) once the SNR exceeds a code-dependent threshold (about $-2$ dB for either BCH code and $-6$ dB for the polar code). Activating the UHF suppresses the leakage by up to $0.87\times$k bit. The leakage suppression has a larger effect on the higher rate \gls{bch} codes compared to the low-rate polar code. Hence, universal hashing markedly widens the SNR safety margin over which the eavesdropper obtains less information. In Fig.~\ref{fig:mi_vs_ber} we compare $\hat I_{\text{proj}}(M_s;Z^n)/k$ versus the \gls{ber} for the same three linear codes with and without \gls{uhf}. This comparison shows the tradeoff between reliability and secrecy, which can be improved when the hash function is activated.


}
\section{Conclusion}
{
In this work, we introduced the \gls{cnbmm} architecture—a neural \gls{mi} estimator designed specifically for \gls{pls}. By using the \gls{vclub} framework and modeling the conditional distribution as a Bernoulli mixture of experts, \gls{cnbmm} improves leakage estimation and avoids the scalability problems of \gls{mine}, which tends to be unstable for high-dimensional data.  
Using this estimator, we also proposed a closed-loop method to design hash functions efficiently, without needing to calculate the smooth min-entropy through exhaustive search. Simulation results show that \gls{cnbmm} provides more accurate and stable leakage estimates for large blocklengths, compared to \gls{mine} or \gls{club}. Overall, our \gls{cnbmm}-based leakage estimator and adaptive hash function design offer a practical tool to bring \gls{pls} for wiretap channels into the finite blocklength setting.

One limitation of this work is that we did not fully explore the large space of parameter combinations~$(n, k, b)$, especially when paired with different types of error correcting codes. A more systematic study would be needed to cover this space, which we leave for future work. Extending \gls{cnbmm} to larger models, or jointly optimizing the leakage estimator, \gls{uhf}, and \gls{ecc}, are promising directions for further research.

}

\bibliographystyle{IEEEtran}
\bibliography{references00,ref2}

\appendix
\section*{Proof of Lemma \ref{Lemma_Hmin_H_gap}}
\addcontentsline{toc}{section}{Appendix: Proof of Lemma \ref{Lemma_Hmin_H_gap}}
{
\begin{proof}
Fix $r>1$ and define $
  \mathcal A_{r}:=
     \bigl\{\, {z}^{n}\in\mathbb R^{n}\;:\;
             | {z}_i|\le r \;\text{for all } i\bigr\}.$
For each $i$, since the received signal at Eve is \( {Z}_i = X_i+N_i,\,\mathds{E}[ {Z}_i]=1\), we have
\(
  \Pr(| {z}_i|>r) = Q\!\bigl((r-1)/\sigma\bigr).
\)
Define the $
  \varepsilon_{r}
  := \Pr\bigl[\mathcal A_{r}^{\mathrm c}\bigr]
  \;\le\;
  2n\,Q\!\bigl((r-1)/\sigma\bigr)$ as the tail probability and we select $r$ such that $\varepsilon_{r}=\varepsilon$. To simplify the notation, we let $\tilde{z}:= {z}^n$ and fix \( \tilde{z}\in\mathcal E\). Then inside $\mathcal A_{r}$, \(v_{ \tilde{z}}=|\supp X^{n}\big|_{ \tilde{z}}|\le 2^{n}\) and \(t_{ \tilde{z}}\le 2^{-n}(2\pi\sigma^{2})^{-n/2}\). Let \(m:=v_{ \tilde{z}}\), \(t:=t_{ \tilde{z}}\). Then conditions~\eqref{EQ_Def_Vz_Tz} hold. Since Shannon's entropy is Schur-concave, and
$   q=(t,\tfrac{1-t}{m-1},\ldots,\tfrac{1-t}{m-1})$ majorizes all $p$, the posterior \gls{pmf}, then $q$ maximizes the entropy. Hence
\begin{align}\label{EQ_schur_convex}
    H(p)\le H(q)=H_{\mathrm b}(t)+(1-t)\log_2(m-1).    
\end{align}

With \(H_{\min}(p)=-\log_{2}t\) and recall $\psi_m(t) \;:=\; H_b(t) + (1-t)\log_2(m-1) + \log_2 t$, we can get the following:
\begin{align}
  -H_{\min}(p)&=\log_2\frac1t                                             \label{eq:1a}\\
   &=\psi_{m}(t)-H_{\mathrm b}(t)-(1-t)\log_{2}(m-1)                        \notag\\
  &\le -H(p)+\psi_{m}(t). \label{EQ_Hmin_UB_p}
\end{align}
  where  
\eqref{EQ_Hmin_UB_p} uses $H(p)\le H_b(t)+(1-t)\log_2(m-1)$ from \eqref{EQ_schur_convex}.  
From the definition
\begin{align}
  H(X\!\mid  \tilde{Z},\mathcal E)
            &:=\mathds E[H(p_{\tilde{z}})\mid  \tilde{Z}\in\mathcal E],          \\
  H_{\min}(X\!\mid  \tilde{Z},\mathcal E)
            &:=\mathds E[H_{\min}(p_{\tilde{Z}})\mid  \tilde{Z}\in\mathcal E],
\end{align}
after taking expectations of \eqref{EQ_Hmin_UB_p}, we have:
    \begin{align}
    -H_{\min}(X\mid  \tilde{Z},\mathcal E)&=        -\mathds E[H_{\min}(p_{\tilde{Z}})\mid\mathcal E]\\
         &\leq -\mathds E[H(p_{\tilde{Z}})\mid\mathcal E]
              +\mathbb{E}\bigl[\psi_{M}(T)\bigr]\label{eq:2a}\\
     &= -H(X\mid  \tilde{Z},\mathcal E)+\mathbb{E}\bigl[\psi_{M}(T)\bigr].          \label{EQ_Hmin_H}
   \end{align}

Now we expand $
P_{X \tilde{Z}}= (1-\varepsilon)P_{X \tilde{Z}}^{(\mathcal E)}
        \;+\;
        \varepsilon\,P_{X \tilde{Z}}^{(\mathcal E^{\mathrm c})},$
where 
\(
   P_{X \tilde{Z}}^{(\mathcal E)}(\,\cdot\,):=
      \dfrac{P_{X \tilde{Z}}(x, \tilde{z})\,\mathbf 1_{\{ \tilde{z}\in\mathcal E\}}}{1-\varepsilon},
\quad
   P_{X \tilde{Z}}^{(\mathcal E^{\mathrm c})}(\,\cdot\,)
      :=\dfrac{P_{X \tilde{Z}}(x, \tilde{z})\,\mathbf 1_{\{ \tilde{z}\notin\mathcal E\}}}{\varepsilon}.
\)

For any measurable $(x, \tilde{z})\in\mathcal X\times\mathcal E$ with $ \tilde{z}\in\mathcal E$,
\begin{align}
P_{X\mid  \tilde{Z}}(x\mid  \tilde{z})
   &=\frac{P_{X \tilde{Z}}(x, \tilde{z})}{P_{\tilde{Z}}( \tilde{z})}
     \\
   &=\frac{(1-\varepsilon)\,P_{X \tilde{Z}}^{(\mathcal E)}(x, \tilde{z})
            +\varepsilon\,P_{X \tilde{Z}}^{(\mathcal E^{\mathrm c})}(x, \tilde{z})}
           {(1-\varepsilon)\,P_{ \tilde{Z}}^{(\mathcal E)}( \tilde{z})
            +\varepsilon\,P_{ \tilde{Z}}^{(\mathcal E^{\mathrm c})}( \tilde{z})}  \\
   &=\frac{(1-\varepsilon)\,P_{X \tilde{Z}}^{(\mathcal E)}(x, \tilde{z})}
           {(1-\varepsilon)\,P_{ \tilde{Z}}^{(\mathcal E)}( \tilde{z})}\\
   &=\frac{1}{1-\varepsilon}\,P_{X\mid  \tilde{Z},\mathcal E}(x\mid  \tilde{z}),\label{EQ_cond_P_smooth_cond_P}
\end{align}
where the third equality is because $P_{ \tilde{Z}}^{(\mathcal E^{\mathrm c})}( \tilde{z})=0\text{ for } \tilde{z}\in\mathcal E\ $. Therefore, after substituting \eqref{EQ_cond_P_smooth_cond_P} into min entropy by taking \(\max_x\) and \(-\log_2\), we have
\begin{align}
   H_{\min}(X\mid  \tilde{Z})
   \ge H_{\min}(X\mid  \tilde{Z},\mathcal E)-\log_2\!\frac1{1-\varepsilon}. \label{EQ_Hmin_Hmin}
\end{align}

Now we want to show that
\begin{align}
  H(X\mid  \tilde{Z},\mathcal E)
  \ge H(X\mid  \tilde{Z})-\frac{\varepsilon}{1-\varepsilon}\,H_{\max}(X).  
\end{align}
  
By convexity of entropy with the expansion \(P_{X \tilde{Z}}=(1-\varepsilon)P^{(\mathcal E)}_{X \tilde{Z}}
          +\varepsilon P^{(\mathcal E^{\mathrm c})}_{X \tilde{Z}}\), we have 
\begin{align}
H(X\mid  \tilde{Z})
&=\sum_{x, \tilde{z}}P_{X \tilde{Z}}(x, \tilde{z})
        \log_2\frac{1}{P_{X\mid  \tilde{Z}}(x\mid  \tilde{z})}                     \notag\\
   &=(1-\varepsilon)\!
        \sum_{x, \tilde{z}}P_{X \tilde{Z}}^{(\mathcal E)}(x, \tilde{z})
        \log_2\frac{1}{P_{X\mid  \tilde{Z},\mathcal E}(x\mid  \tilde{z})}          \notag\\
   &\quad
     +\varepsilon\!
        \sum_{x, \tilde{z}}P_{X \tilde{Z}}^{(\mathcal E^{\mathrm c})}(x, \tilde{z})
        \log_2\frac{1}{P_{X\mid  \tilde{Z},\mathcal E^{\mathrm c}}(x\mid  \tilde{z})}\notag\\
   &= (1-\varepsilon)H(X\mid  \tilde{Z},\mathcal E)
   +\varepsilon H(X\mid  \tilde{Z},\mathcal E^{\mathrm c})\\
&\le (1-\varepsilon)H(X\mid  \tilde{Z},\mathcal E)
    +\varepsilon H_{\max}(X).
\end{align}

After rearrangement, we have
\begin{align}
H(X\mid  \tilde{Z},\mathcal E)
  &\ge \frac{H(X\mid  \tilde{Z})}{1-\varepsilon}-\frac{\varepsilon}{1-\varepsilon}\,H_{\max}(X)\\
  &\ge H(X\mid  \tilde{Z})-\frac{\varepsilon}{1-\varepsilon}\,H_{\max}(X).\label{EQ_cond_H_cond_H}
\end{align}

Recall the definition of smooth min entropy in \eqref{eq:def_HminSmooth}. Then we can choose
\(  P_{X \tilde{Z}}:=P_{X \tilde{Z}}\bigl[\cdot\cap\{ \tilde{Z}\in\mathcal E\}\bigr]\)
in \eqref{eq:def_HminSmooth} results in 
\begin{align}
   H_{\min}^{\varepsilon}(X\mid  \tilde{Z})
   \ge H_{\min}(X\mid  \tilde{Z};  P_{X \tilde{Z}})
   = H_{\min}(X\mid  \tilde{Z},\mathcal E).                          \label{EQ_smoth_H_min_H}
\end{align}

Combining \eqref{EQ_smoth_H_min_H}, \eqref{EQ_Hmin_Hmin},  \eqref{EQ_Hmin_H},  and \eqref{EQ_cond_H_cond_H}, we complete the proof.
\end{proof}
}

\end{document}